\documentclass[onecolumn,draftcls]{IEEEtran}
\usepackage{lineno,hyperref}
\usepackage{tabularx}
\usepackage{amsmath}
\usepackage{amssymb}
\usepackage{multirow}
\usepackage[table,xcdraw]{xcolor}
\usepackage[linesnumbered,ruled,vlined]{algorithm2e}
\usepackage{color}
\usepackage{subfigure}
\usepackage{graphicx}

\ifCLASSINFOpdf
\else
\fi
\begin{document}
%
\title{Ensemble of meta-heuristic and exact algorithm based on the divide and conquer framework for multi-satellite observation scheduling}
%
%
%

\author{Guohua~Wu,~\IEEEmembership{Member,~IEEE,}
        Qizhang~Luo,
        Xiao~Du,
        Yingguo~Chen,
        Xinwei~Wang,
        Ponnuthurai~Nagaratnam~Suganthan,~\IEEEmembership{Fellow,~IEEE}
\thanks{This work has been submitted to the IEEE for possible publication. Copyright may be transferred without notice, after which this version may no longer be accessible.}
\thanks{Guohua Wu and Xiao Du are with the School of Traffic and Transportation Engineering, Central South University, Changsha 410075, Hunan, China. (E-mails: guohuawu@csu.edu.cn, 1272068435@qq.com)}
\thanks{Qizhang Luo is with the School of Traffic and Transportation Engineering, Central South University, Changsha 410075, Hunan, China, also with the Department of Electrical and Computer Engineering, National University of Singapore, 119260, Singapore. (E-mails: qz\_luo@csu.edu.cn)}
\thanks{Yingguo~Chen is with the College of Systems Engineering, National University of Defense Technology, Changsha 410073, Hunan, China. (E-mail: ygchen@nudt.edu.cn)}
\thanks{Xinwei Wang (\emph{Corresponding author}) is with the School of Traffic and Transportation Engineering, Central South University, Changsha 410075, Hunan, China, also with the Department of Transport and Planning, Stevinweg 1, 2628 CN Delft, the Netherlands. (E-mails: xinwei.wang.china@gmail.com)}
\thanks{Ponnuthurai~Nagaratnam~Suganthan is with the School of Electrical and Electronic Engineering, Nanyang Technological University, 639798, Singapore. (Email:epnsugan@ntu.edu.sg
 )}}

\maketitle

\begin{abstract}
    Satellite observation scheduling plays a significant role in improving the efficiency of Earth observation systems. To solve the large-scale multi-satellite observation scheduling problem, this paper proposes an ensemble of meta-heuristic and exact algorithm based on a divide-and-conquer framework (EHE-DCF), including a task allocation phase and a task scheduling phase. In the task allocation phase, each task is allocated to a proper orbit based on a meta-heuristic incorporated with a probabilistic selection and a tabu mechanism derived from ant colony optimization  and tabu search  respectively. In the task scheduling phase, we construct a task scheduling model for every single orbit, and use an exact method (i.e., branch and bound, B\&B) to solve this model. The task allocation and task scheduling phases are performed iteratively to obtain a promising solution. To validate the performance of EHE-DCF, we compare it with B\&B, three divide-and-conquer based meta-heuristics, and a state-of-the-art meta-heuristic. Experimental results show that EHE-DCF can obtain higher scheduling profits and complete more tasks compared with existing algorithms. EHE-DCF is especially efficient for large-scale satellite observation scheduling problems.
\end{abstract}


%
\IEEEpeerreviewmaketitle

\section{Introduction}

Earth observation satellites (EOSs) are widely used in sensing Earth's surface and surrounding atmosphere. The extremely useful imaging capabilities of EOSs have played an important role in resource exploration, disaster surveillance, urban planning, and environmental monitoring \cite{Wu2016coordinated,li2020auto}. In recent years, although the number of EOSs is increasing continuously and has reached 906 by January 1st 2021, the satellites are still insufficient for serving numerous Earth observation requests \cite{Wang2020robust}. Therefore, the EOS scheduling problem that aims to accomplish as many observation requests as possible forms an essential component in the Earth observation satellite systems.

\begin{figure}[htb]
	\centering
	\includegraphics[width=0.5\textwidth,trim=3 3 3 3,clip]{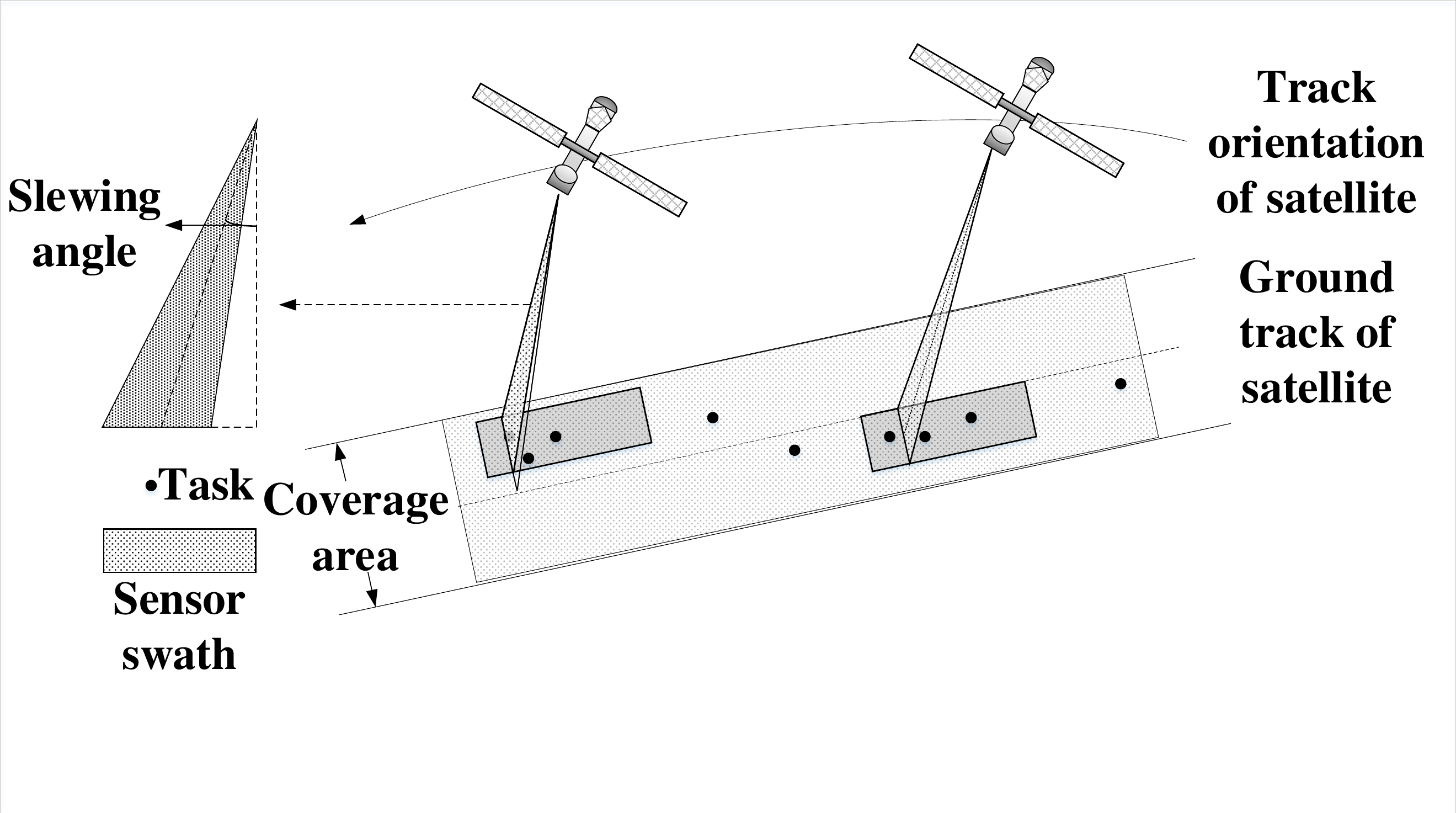}
	\caption{Satellite imaging activity.}
	\label{Fig:1}
\end{figure}

An illustration of the EOS imaging activity is shown in Figure \ref{Fig:1}. An EOS flies around the Earth, and its sensor could generate an observation strip with a certain width and length when passing over a target. To observe multiple targets, the EOS needs to conduct certain operations for the transfer between two consecutive observation tasks, such as attitude slewing and stabilization. Besides, the EOS can only perform imaging operations within a limited time window~\cite{HU201974, tobias2021agile, wang2019onboard} when it can pass a ground target and catch sight of the target. 
During the observation process, each ground target has different visible time windows for different EOSs. Meanwhile, an EOS can observe the same target on different orbits, thus there may be multiple visible time windows between an EOS and a target. Therefore, the scheduling problem of an EOS belongs to a class of single-machine scheduling problems with time window constraints and sequence-dependent setup time, which has been proved to be NP-hard \cite{waiming2019two, liao2007imaging}.

Although many impressive studies have been carried out to address EOS scheduling problems \cite{liao2007imaging, PENG2020104946, huang2021revising, ZHU2010220, WANG2016125}, the increasing number of orbiting satellites and user demands has posed new challenges on multi-satellite scheduling problems with large-scale tasks in practical applications. In this paper, we aim to tackle a multi-satellite, multi-orbit, and large-scale Earth observation scheduling problem. The difficulties for solving this kind of problem can be viewed in two aspects. First, the number of candidate EOSs and visible time windows for serving a task, as well as the number of tasks could increase the problem complexity exponentially. Second, multiple satellites indicate that the constraints would be more complex to guarantee the collaboration of satellites, which increases the difficulties for solving the problem compared with single satellite scheduling problems. In regard to these difficulties, the exact algorithms developed in existing literature are no longer suitable for large-scale scheduling problems, as the computational time of exact algorithms is unacceptable \cite{wang2019exactheuristic, chen2018exact}. On the other hand, although heuristic and meta-heuristics have been widely used to solve large-scale scheduling problems \cite{ZHANG2019100560, CHEN2021100912, DU2019100576, waiming2019two}, the scheduling results of these methods may not have performance guarantees. Therefore, it could be natural and of great significance to develop efficient satellite scheduling methods combing the exact and meta-heuristics, which can bring together the advantages of these two types of optimization algorithms to achieve a higher scheduling performance, to satisfy more user demands and improve the efficiency of the Earth observation systems.

Motivated by this, in this study, we develop a scheduling framework based on the well-known divide-and-conquer principle \cite{abel1990}, which decomposes a large-scale EOS scheduling problem into multiple sub-problems to reduce the complexity of solving the problem. We treat the orbits of satellites as the resources providing imaging services and propose a novel scheduling method under the divide-and-conquer framework (DCF). This method comprises two phases: task allocation among multiple orbits and task scheduling on a single orbit. In the task allocation phase, we develop a meta-heuristic allocation method based on the idea of pheromone in ant colony optimization (ACO) and the tabu mechanism in tabu search (TS). After the task allocation phase, multiple sub-problems that schedule tasks on each orbit are generated. Afterward, in the task scheduling phase, we construct an integer programming model for each single orbit scheduling problem and utilize a branch and bound (B\&B) method to solve this model exactly. The task allocation and single orbit scheduling in the two phases are executed iteratively and interactively until a promising solution is obtained.

The overall approach can be viewed as an ensemble of meta-heuristic and exact algorithm based on the DCF (EHE-DCF). It provides a new paradigm that cooperatively uses meta-heuristics and exact mathematical programming approaches to solve complex and large-scale combinatorial optimization problems. When confronting a complex and large-scale optimization problem, meta-heuristics may not be effective to find a high-quality solution, while exact methods generally cannot obtain an optimal solution with affordable computation time. EHE-DCF can partition the original problem into multiple sub-problems by using a meta-heuristic and then solves each simpler sub-problem via exact and mature mathematical programming approaches. The obtained solution would be more high-quality, while the optimization process would be more efficient.

The main contributions of this paper are summarized as follows: 

(1) We propose an ensemble of meta-heuristic and exact algorithm based on the DCF (EHE-DCF) to address a multi-satellite, multi-orbit, and large-scale Earth observation scheduling problem. The proposed framework is a new paradigm that decomposes a large-scale scheduling problem into sub-problems and combines the advantages of meta-heuristic and exact algorithm.

(2) In EHE-DCF, we treat the orbits of satellites as independent resources that could provide imaging services, and divide the scheduling process into two phases, i.e., task allocation among multiple orbits in the first phase and task scheduling on every single orbit in the second phase. These two phases are iteratively and interactively performed to further improve the quality of the solution. 

(3) We design a meta-heuristic based on the pheromone mechanism used in ACO and the tabu mechanism used in TS to realize the effective task allocation in the first phase of EHE-DCF. In addition, we employ a mathematical model and a corresponding B\&B method to address the task scheduling problem on every single orbit in the second phase.

(4) Extensive experiments on EOS scheduling instances with multi-satellite, multi-orbit, and large-scale tasks are conducted to validate the performance of EHE-DCF. Specifically, EHE-DCF is compared with five existing approaches: an exact method without the DCF (denote as pure B\&B), three DCF based meta-heuristics, i.e., greedy algorithm based on DCF (GR-DCF), simulated annealing algorithm based on greedy neighborhood structure and DCF (SANS1-DCF), and simulated annealing algorithm based on random neighborhood structure and DCF (SANS2-DCF), as well as a sate-of-the-art meta-heuristic (ASA) \cite{WU2017576}. The experimental results demonstrate the superiority of EHE-DCF.

The rest of the paper is structured as follows. Section \ref{s2} surveys the related works. Section \ref{s3} provides the scheduling framework based on the divide-and-conquer principle. In Section \ref{s4}, we present a mathematical model for the single orbit task scheduling. We introduce the EHE-DCF in Section \ref{s5}. The simulation experiments and the results are detailed in Section \ref{s6}. Finally, Section \ref{s7} concludes the paper.

\section{Related works}\label{s2}
At present, considerable achievements have been made in the domain of EOS scheduling problems. The models established by scholars include mathematical programming models \cite{Wang2018fixed, LI20193258, wang2020expectation,Wang2020robust}, constraint satisfaction problem models \cite{deng2020twophase, sun2008Satellites, plaunt1999satellite}, knapsack problems \cite{luo2020hybrid, vasquez2003upper,WANG20191011}, graph-based problems \cite{WANG2016125, wang2019time, SARKHEYLI2013287}. The algorithms for EOS scheduling problems can be roughly classified into exact, heuristic, and meta-heuristic approaches.

In general, exact algorithms are feasible in tackling small-scale EOS scheduling problems \cite{HU201974}. For instance, Gabrel and Vanderpooten \cite{GABREL2002533} formulated the EOS scheduling problem as the selection of a multiple criteria path in a graph without a circuit. This problem was solved by the generation of efficient paths and the selection of a satisfactory path using multiple criteria interactive procedure. Hu et al. \cite{HU201974} conducted a study on the application of exact algorithms to EOS constellations and proposed a branch-and-price algorithm to solve the EOS constellation imaging and downloading integrated scheduling problem. Peng et al. \cite{PENG2020104946} investigated the agile satellite scheduling problem with time-dependent profits and solved the problem based on an adaptive-directional dynamic programming algorithm with decremental state space relaxation.

Exact algorithms can get optimal scheduling results, but for NP-hard optimization problems, the required computation time of exact algorithms usually increases exponentially with the increase of problem scale. Thus, heuristics and meta-heuristics are carried out to solve the EOS scheduling problems. For Example, Wu et al.~\cite{WU2017576} developed a formal model for EOS scheduling problems, and presented an adaptive SA–based scheduling algorithm integrated with a dynamic task clustering strategy. Huang et al.~\cite{huang2009simulation} presented a multi-objective chance constrained programming model for electronic reconnaissance satellites scheduling problem, and proposed a Monte Carlo simulation based multi-objective evolutionary algorithm. Many scholars have applied the ACO algorithm to solve EOS scheduling problems \cite{gao2013multi, gao2013ant, ZHANG20142816, zhang2018ant}. For example, Gao et al. \cite{gao2013multi} constructed an acyclic directed graph model for the EOS scheduling problem and presented a novel hybrid ACO method. Zhang et al. \cite{ZHANG20142816, zhang2018ant} presented a complex independent set model for multi-satellite control resource scheduling problem, and proposed an ACO based algorithm, in which the pheromone trail is updated by two stages to avoid local optima. Wang et al. \cite{wang2009scheduling} established an integer programming model for the EOS scheduling problem, and proposed a hybrid ACO algorithm, where the pheromone was used to indicate how to choose the request to schedule. These existing works inspired us to design a meta-heuristic based on the pheromone mechanism of ACO to realize the task allocation in the first phase.

In recent years, a new trend for solving the EOS scheduling problem is to decompose the large-scale scheduling problem into several small-scale scheduling problems that can be solved separately. For instance, Xu et al. \cite{XU202093} transfered the very large area observation problem into a set covering problem with constraints and solved the problems based on a three-phase algorithm. Liu et al. \cite{XIAOLU2014687} decomposed the scheduling problem into two sub-problems: task assignment and task merging. Our study is distinguished from these studies in two aspects. First, we decompose the scheduling problem into sub-problems based on a divide-and-conquer framework, which can solve the problem iteratively and interactively. Second, we propose an ensemble of meta-heuristic and exact algorithm, which combines the advantages of these two kinds of algorithms, thereby improving the efficiency of the optimization process significantly.

Furthermore, in the above studies, scholars usually formulated satellites as resources and assumed that each task has at most one observation time window on each resource. However, a satellite may have multiple orbits to provide multiple observation time windows. If the scheduling period is long enough, a satellite will pass over a target multiple times. Hence, the observation time windows for a task on a satellite will not be unique, which makes the EOS scheduling problem more difficult. A better solution to tackle this issue was proposed in Wu et al.~\cite{wu2012multisatellite}, which formulated the orbits of satellites as resources such that there will be at most one observation window for each task on each resource, making the EOS scheduling problem easier to model. In this paper, we also formulate the orbits of the satellites as observation resources.

\section{Divide-and-conquer-based scheduling framework}\label{s3}

\begin{figure}[htb]
	\centering
	\includegraphics[width=\linewidth]{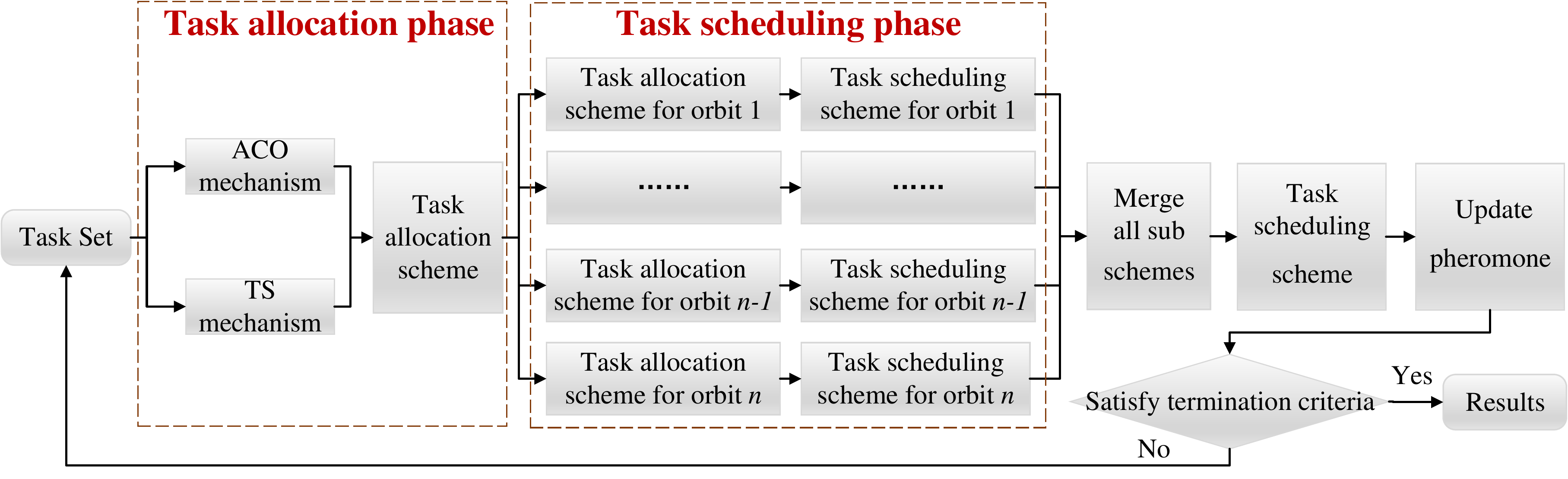}
	\caption{The scheduling framework based on divide-and-conquer principle.}
	\label{Fig:2}
\end{figure}

As Figure \ref{Fig:1} shows, an EOS could generate an observation strip of a certain width and length when passing over a target. The width and length of the strip are determined by the altitude of the satellite, as well as the view field, the slewing angle, and the observation duration of the sensor \cite{wang2019onboard,wei2021scheduling}. In order to facilitate modeling, we assume that all observation targets are point targets and a target is termed a task. Besides, the orbits of the satellites are termed as resources that could provide imaging services. To schedule the satellite resources efficiently, we propose a novel scheduling framework based on the divide-and-conquer principle (DCF). The framework comprises two iterative phases: task allocation phase among multiple orbits and task scheduling phase on every single orbit, whose workflow is shown in Figure \ref{Fig:2}.

In the task allocation phase, we develop a meta-heuristic to allocate tasks to orbits. We first calculate the probability of each task to be allocated to each orbit. The calculation of the allocation probability between a task and an orbit is inspired by the idea of the pheromone mechanism of ACO \cite{ zhang2018ant}. Once a task is allocated to an orbit, the intensity of pheromone between the task and the orbit will be increased. Besides, the tabu mechanism of TS is adopted in this phase to avoid premature convergence. Further details of the meta-heuristic are provided in Section \ref{s5}.

In the task scheduling phase, a set of task scheduling sub-problems on every single orbit is separately solved by a B\&B method based on the allocation results in the task allocation phase. The scheduling scheme of an orbit is exactly regarded as the solution of a sub-problem, and the overall scheduling scheme can be obtained by merging all the sub-scheduling schemes.

The final scheduling results are obtained through iteratively performing the task allocation phase and task scheduling phase until the algorithm termination conditions are met. These two phases are interconnected via the pheromone mechanism and tabu mechanism. This framework can effectively reduce the scheduling complexity of the original problem and obtain a more promising solution by merging a set of sub-solutions obtained by a B\&B method that solves a set of relatively small-scale sub-problems.

\section{Mathematical model for task scheduling on an orbit}\label{s4}
In this section, we develop a task scheduling model for a single orbit. Satellite observation operations in practical applications are affected by various factors, such as cloud coverage, imaging data transmission, and satellite malfunction \cite{Wang2020robust}. For the convenience of modeling, we assume that the impacts of these real-world factors are ignored. Besides, we assume that each task is desired to be observed once, without repeated observation requests. The task scheduling model aims to maximize the overall profits of all the scheduled tasks, while satisfying the constraints related to satellite operations, including satellite transfer time between two consecutive tasks, energy capacity, and memory capacity. The profit of a task represents the importance and value to the user of completing the observation task \cite{wang2020expectation,huang2009simulation,wang2009scheduling}.

The used notations are summarized in Table \ref{table:1}. Let $O = \left\{{{O_1},{O_2},\cdots,{O_H}} \right\}$ be the set of orbits within the predefined scheduling horizon and $H$ the number of orbits. Denote $T = \left\{ {1,2, \cdots,N} \right\}$ as the set of tasks and $N$ the number of tasks. Define $T_k \in T$ as the set of tasks allocated to orbit $k \in O$. Each orbit $k$ is associated with a memory capacity ${M_k}$ and an energy capacity ${E_k}$. The observation activity consumes energy and memory resources on each orbit. Define ${e_k}$ and ${m_k}$ as the energy consumption and the memory consumption for a task on orbit $k$. Each task $i \in {T_k}$ is endowed with an observation profit ${\omega _i}$, a slewing angle ${\theta _{ik}}$, and a time window [$w{s_{ik}}$,$w{e_{ik}}$] specified by its earliest possible observation time $w{s_{ik}}$ and its latest possible observation time $w{e_{ik}}$.

\begin{table}[htb]
	\caption{Notations for the scheduling model.}\label{table:1}
	\centering
	\begin{tabular}{lp{0.7\linewidth}}
	\hline
	Notations & Description \\
	\hline
    $O$ &Set of orbits\\
    $H$ &Number of orbits\\
    $T$ &Set of tasks\\
    $N$ &Number of tasks\\
    $T_k$ &Set of tasks on orbit $k$, $T_k \in T$\\  
    $x_{ij}^k$ &Decision variables\\
    $\omega_i$ &Profit of task $i$\\
    $M_k$, $E_k$ &Memory capacity and energy capacity of orbit $k$\\
    $m_k$, $e_k$ &Memory and energy consumption for each time unit of observation of orbit $k$\\
    $[ws_{ik}, we_{ik}]$&Time window for task $i$ on orbit $k$\\
    $\theta_{ik}$ &Slewing angle for task $i$ on orbit $k$\\
    $st_{ij}^k$ &Transfer time between task $i$ and task $j$ on orbit $k$\\
    $td_k$, $tu_k$, $ts_k$ &Times of sensor shutdown, startup and attitude stabilization on orbit $k$\\
    $v$ &Slewing velocity of sensors\\
	\hline
	\end{tabular}
\end{table}

Satellite transfer time is required to observe two different tasks successively. Specifically, after observing a task $i \in {T_k}$, the satellite needs a sequence of transformation operations to observe the next task $j \in {T_k}$, including sensor shutdown, slewing, attitude stability, and startup. Denote $t{u_k}$, $t{d_k}$, and $t{s_k}$ as the time consumption of sensor startup, shutdown, and attitude stability on the orbit $k$, respectively. Let $v$ be the slewing velocity of the satellite. The transfer time $st_{ij}^k$ can be computed as
\begin{equation}\label{eq1}
    st_{ij}^k = t{d_k} + \left| {{\theta _{ik}} - {\theta _{jk}}} \right|/v + t{s_k} + t{u_k}.
\end{equation}

To accomplish the model, we introduce  binary decision variables $x_{ij}^k$, and define $x_{ij}^k = 1$ if 
task $i$ is the immediate predecessor of the task $j$ on orbit $k$; otherwise $x_{ij}^k = 0$. Note that there exist two dummy tasks used to start or terminate sensors when the task index equals 0 or $N + 1$. The dummy tasks do not have real profit (i.e., ${\omega _0} = 0$, ${\omega _{N + 1}} = 0$). Thus, the integer programming formulation of the task scheduling on the single orbit is constructed as
\begin{equation}\label{eq2}
    \max \sum_{i \in T_k}\sum_{j \in T_k \cup \{N+1\},\\j \ne i}x_{ij}^k \cdot \omega_i,
\end{equation}
\begin{equation}\label{eq3}
    \sum_{i \in T_k}\sum_{j \in T_k \cup \{N+1\},\\j \ne i}x_{ij}^k \cdot e_i \le E_k,
\end{equation}
\begin{equation}\label{eq4}
    \sum_{i \in T_k}\sum_{j \in T_k \cup \{N+1\},\\j \ne i}x_{ij}^k \cdot m_k \cdot (we_{ik} - ws_{ik}) \le M_k,
\end{equation}
\begin{equation}\label{eq5}
    x_{ij}^k \cdot (ws_{ik} - we_{ik} - st_{ij}^k) \ge 0, \forall i,j \in T_k,
\end{equation}
\begin{equation}\label{eq6}
    \sum_{j \in T_k \cup \{0\},\\j \ne i}x_{ij}^k \le 1, \forall i \in T_k,
\end{equation}
\begin{equation}\label{eq7}
    \sum_{j \in T_k \cup \{N+1\},\\j \ne i}x_{ij}^k \le 1, \forall i \in T_k,
\end{equation}
\begin{equation}\label{eq8}
    \sum_{j \in T_k \cup \{0\},\\j \ne i}x_{ij}^k -\sum_{j \in T_k \cup \{N+1\},\\j\neq i }x_{ij}^k \neq 0, \forall i \in T_k,
\end{equation}
\begin{equation}\label{eq9}
x_{0,N+1}^k + x_{N+1,0}^k = 0,
\end{equation}
\begin{equation}\label{eq10}
    \sum_{j\in T_k}x_{0,j}^k = 1,
\end{equation}
\begin{equation}\label{eq11}
    \sum_{i\in T_k}x_{i,N+1}^k = 1.
\end{equation}

The objective function (\ref{eq2}) aims to maximize the entire observation profits of all the scheduled tasks. Constraints (\ref{eq3}), (\ref{eq4}), and (\ref{eq5}) represent the energy, memory, and time window constraints, respectively. Constraints (\ref{eq6}) and (\ref{eq7}) indicate that there is at most one predecessor task and one subsequent task for each real task. Constraint (\ref{eq8}) guaranteed that the number of predecessor tasks and the number of subsequent tasks for a real task are equal. Moreover, constraints (\ref{eq6}), (\ref{eq7}) and (\ref{eq8}) also show that each task can be observed at most once. Constraint (\ref{eq9}) indicates that the dummy tasks cannot be used as adjacent tasks. Since adjacent dummy tasks do not contribute to the objective function, we force $x_{0,N + 1}^k{\rm{ = }}0$ and $x_{N + 1,0}^k = 0$. Constraints (\ref{eq10}) and (\ref{eq11}) ensure that there must be a real task after the dummy task 0 and a real task before the dummy task $N+1$. Similar EOS scheduling models can be found in \cite{WU20131884,WU2017576,WANG20161}, in which constraints (\ref{eq9})–(\ref{eq11}) constructed in this paper are not considered.

\section{Ensemble of meta-heuristic and exact algorithm}\label{s5}
\subsection{Meta-heuristic for task allocation}

To cooperate with the proposed DCF, we propose a novel meta-heuristic method hybriding the tabu mechanism of TS and the pheromone mechanism of ACO, for the task allocation phase. In detail, the tabu mechanism is utilized to modify the orbit set for task allocation, while the pheromone mechanism is adopted to select appropriate orbits for tasks. Furthermore, we introduce three factors, including heuristic factor, pheromone trail factor, and feedback factors to implement the above two mechanisms.

Before detailing the proposed meta-heuristic method, we clarify some definitions for convenience. Denote $CO_i$ as an orbit set in which all orbits have visible time windows for the task $i$. Although all the orbits in $CO_i$ are visible to $i$, some of them would not be used due to the tabu mechanism. Thus, denote $CO_i'(n) \subseteq CO_i$ as the available orbits at the $n$-th iteration. Let ${w_i}(n)$ be the allocation priority of a task $i$ at the $n$-th iteration. Denote ${\sigma _{ik}}(n) \in [0,l]$ as the tabu factor between a task $i$ and an orbit $k$ at the $n$-th iteration, where $l$ is the tabu step. ${\sigma _{ik}}(n)$ is used to check whether an orbit $k$ can be utilized in $CO_i'(n)$ when allocating $i$. Further, define ${\eta _{ik}}(n)$ and ${\tau _{ik}}(n)$ as the heuristic factor and the pheromone trail factor between a task $i$ and an orbit $k$, respectively. ${\eta _{ik}}(n)$ and ${\tau _{ik}}(n)$ are used to calculate the allocation probability ${p_{ik}}(n)$ of assigning a task $i$ to an orbit $k$ at the $n$-th iteration, which can be expressed by
\begin{equation}\label{eq12}
    p_{ik}(n) = 
    \frac{ \left [\tau _{ik}(n) \right]^\alpha \times \left[ \eta _{ik}(n) \right]^\beta } 
    {\sum_{t \in CO_i'(n)} \left[ \tau _{it}(n) \right]^\alpha \times \left[ \eta _{it}(n)\right]^\beta},
\end{equation}
\noindent where parameters $\alpha $ and $\beta $ are real numbers that determine the relative influence of the pheromone trail and the heuristic information.
\begin{figure}[htb]
	\centering
	\includegraphics[width=\linewidth]{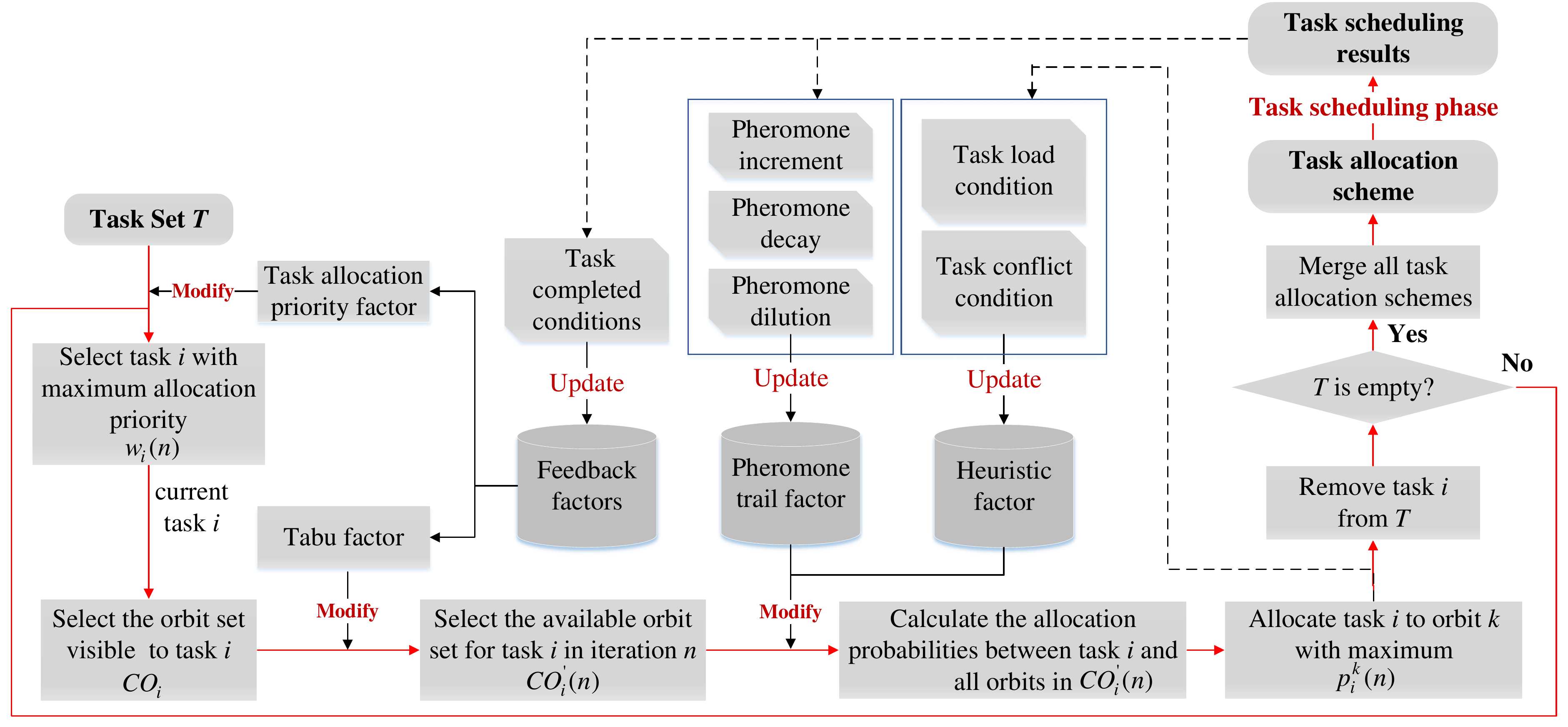}
	\caption{The task allocation process.}
	\label{Fig:3}
\end{figure}

An overview of the meta-heuristic for task allocation is illustrated in Figure \ref{Fig:3}. When performing the task allocation, tasks in the task set $T$ are first sorted in descending order according to their allocation priority ${w_i}(n)$. Then, an available orbit set $CO_i'(n)$ is selected for the task $i$ with the highest allocation priority. Afterwards, the allocation probabilities between the task $i$ and each orbit in $CO_i'(n)$ are calculated, and $i$ is allocated to an orbit $k$ according to the allocation probability ${p_{ik}}(n)$. Meanwhile, the heuristic factor is updated and the assigned task $i$ is removed from $T$. The above processes are repeated iteratively until $T$ is empty. After each iteration, the pheromone trail factor and feedback factors will be updated according to the scheduling results. The results of task allocation after each iteration are different, which can make the algorithm escape from local optima and converge to better solutions gradually.

In particular, the heuristic factor is designed based on the task load condition and task conflict condition, and it will be updated once a task is allocated. The pheromone trail factor is updated at each iteration, during which it will be increased, decayed, and diluted according to the scheduling results. The calculation methods of the heuristic factor, pheromone trail factor, and feedback factors are presented as follows.

(1) \textit{Heuristic factor}. Denote $T_k'$ as a set of tasks already allocated to the orbit $k$, the load degree of $k$ can be expressed by ${\varepsilon _k} = \left| {T_k'} \right|$ (i.e., the number of tasks already allocated to $k$). When allocating a new task $i$, the orbit $k$ in $CO_i'(n)$ presents one of the following two states: (i) There are no allocated tasks on the orbit $k$ (i.e., ${\varepsilon _k} = 0$) and (ii) there are some tasks already allocated to the orbit $k$ (i.e., ${\varepsilon _k} > 0$). In the second state, the task $i$ to be assigned may conflict with other tasks in $T_k'$. We use a binary variable $conf_{is}^k \in \left\{ {0,1} \right\}$ to measure the conflict between the tasks $i$ and $s \in T_k'$. If $conf_{is}^k = 1$, task $i$ conflicts with task $s$; 0 otherwise. The criteria for judging whether two tasks are in conflict is expressed by
\begin{equation}\label{eq13}
    \begin{cases} 
        we_{ik} + |\theta_{ik} - \theta_{sk}|/v > ws_{sk}, & \text{if }ws_{sk} > ws_{ik}, \\
        we_{sk} + |\theta_{ik} - \theta_{sk}|/v > ws_{ik}, & \text{if }ws_{sk} \le ws_{ik}.
    \end{cases}
\end{equation}

Then, let ${\xi _{ik}}$ be the conflict degree between task $i$ and all the tasks in $T_k'$, it can be calculated according to
\begin{equation}
    \xi_{ik} = \sum_{s\in T_{k}'}{conf_{is}^k}.
\end{equation}

\begin{figure}[htb]
	\centering
	\includegraphics[width=0.8\textwidth]{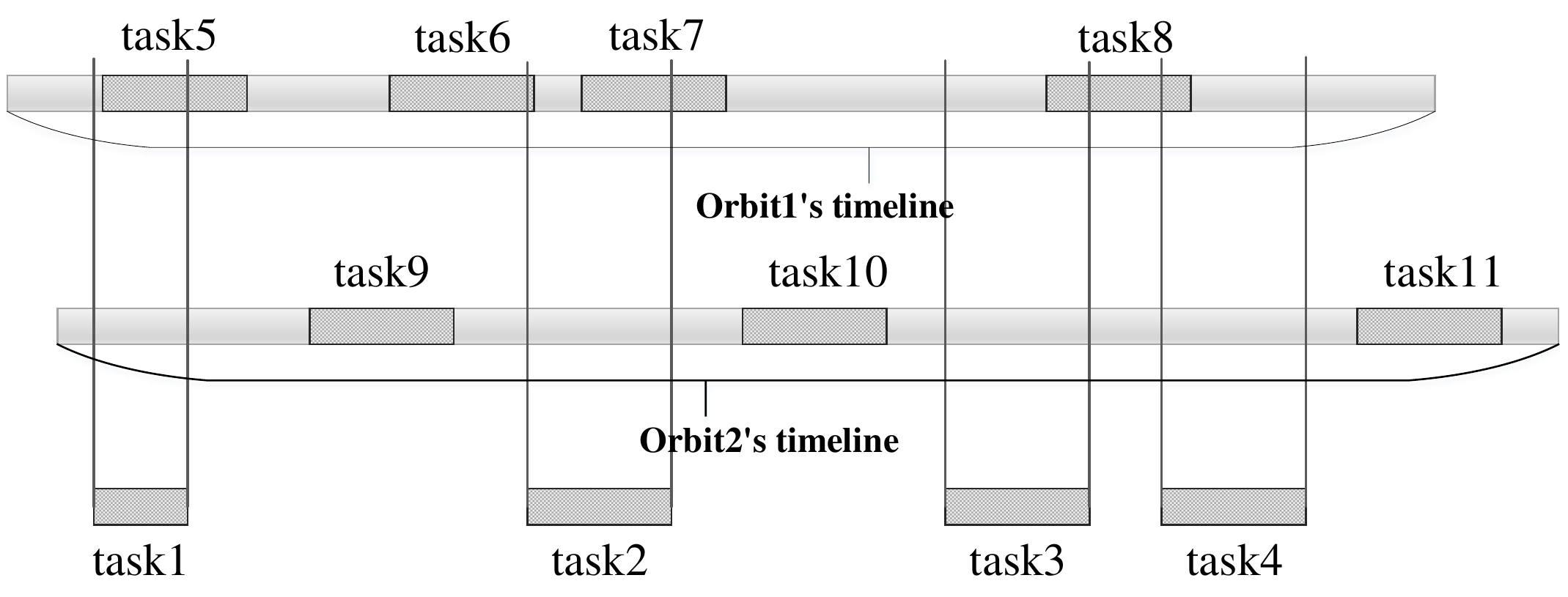}
	\caption{Conflict and Load condition.}
	\label{Fig:4}
\end{figure}

Figure~\ref{Fig:4} shows an example of the conflict degree and load degree, in which $\xi _{11}= 1$, $\xi _{12}= 0$, $\xi _{21}= 2$, $\xi _{22}=0$, $\xi _{31}= 1$, $\xi _{32}= 0$, $\xi _{41}= 1$, $\xi _{42}= 0$, ${\varepsilon _1}{\rm{ = 4}}$, and ${\varepsilon _2}{\rm{ = 3}}$.

Denote ${\bar \varepsilon _k}$ and ${\bar \xi _{ik}}$ as the normalized values of ${\varepsilon _k}$ and ${\xi _{ik}}$ respectively, they can be calculated by
\begin{equation}\label{eq15}
    {\bar \varepsilon _k} = (\sum_{s\in CO_{i}(n)}{\varepsilon_s - \varepsilon_k})/\sum_{s\in CO_{i}(n)}{\varepsilon_s},
\end{equation}
\begin{equation}\label{eq16}
    {\bar \xi _{ik}} = (\sum_{s\in CO_{i}(n)}{\xi_{is} - \xi_{ik}})/\sum_{s\in CO_{i}(n)}{\xi_{is}}.
\end{equation}

Therefore, the value of the heuristic factor $\eta _{ik}(n)$ can be obtained by
\begin{equation}\label{eq17}
    \eta _{ik}(n) = a\cdot{\bar \varepsilon _k} + b\cdot{\bar \xi _{ik}},
\end{equation}
where $a$ and $b$ are the weights of $\bar \varepsilon _k$ and $\bar \xi _{ik}$, respectively.

(2) \textit{Pheromone trail factor}. We use $\rho  \in (0,1)$ and $\lambda $ to represent the pheromone decay parameter and pheromone dilution parameter, respectively. The increment of the pheromone between a task $i$ and an orbit $k$ can be expressed as
\begin{equation}\label{eq18}
    \Delta {\tau _{ik}} = \frac{{\gamma (n)}}{{\lambda  \cdot Num(n)}},
\end{equation}
where $\gamma (n)$ and $Num(n)$ are scheduling profits and number of scheduled tasks at the $n$-th iteration, respectively. Denote ${\gamma ^ * }$ as the current best scheduling profits, it can be calculated as
\begin{equation}\label{eq19}
    {\gamma ^ * } = \max \left\{ {\gamma (1),\gamma (2), \cdots ,\gamma (n)} \right\}.
\end{equation}

When $\sum_{j\in T_k \cup \{N+1\},\\j\neq i}{x_{ij}^k} = 1$, which indicates that the task $i$ is successfully scheduled, the pheromone trail factor ${\tau _{ik}}(n)$ would be updated based on the following formulas
\begin{equation}\label{eq20}
    \begin{cases}
        {\tau _{ik}}(1 + n) = (1 - \rho ) \cdot {\tau _{ik}}(n) + \Delta {\tau _{ik}},\\
        {\tau _{ik}}(1) = 1.
    \end{cases}        
\end{equation}

(3) \textit{Feedback factors}. The feedback factors consist of the allocation priority factor ${w_i}(n)$ and tabu factor ${\sigma _{ik}}(n)$. After the task scheduling on each orbit has been finished, we use ${w_i}(n)$ and ${\sigma _{ik}}(n)$ to update the allocation orders of the task $i$, as well as the elements in $CO_i'(n)$ at the $(n+1)$-th iteration, which will affect the task allocation process at the next iteration.

The initial value of the allocation priority factor is the initial profit of the task $i$ (i.e., ${w_i}(1) = {\omega _i}$). If the task $i$ has not been successfully scheduled (i.e., $\sum_{j\in T_k \cup \{N+1\},j\neq i}{x_{ij}^k} = 0$), ${w_i}(n + 1)$ will be updated; otherwise it keeps unchanged, which can be expressed as
\begin{equation}\label{eq21}
    \begin{cases}
        w_i(n+1) = w_i(n) - c &\text{If }\sum_{j\in T_k \cup \{N+1\},j\neq i}{x_{ij}^k} = 0,\\
        w_i(n+1) = w_i(n) &\text{Otherwise},
    \end{cases}        
\end{equation}
where $c\in \mathbb{Z}_{+}$ is a weight decay parameter. The allocation priority of a task will be decreased if the task is not scheduled, thereby ensuring that the task that has not been scheduled for several times will be adjusted backward in the task allocation sequence.

The tabu factor $\sigma _{ik}(n)$ checks whether an orbit is available or not. The initial value of $\sigma _{ik}(n)$ is 0, and it will be updated iteratively in the later iterations. We propose 7 heuristic rules to determine the value of $\sigma _{ik}(n)$. Specifically, the initialization of the tabu process is shown in \textbf{Rule1}. If ${\sigma _{ik}}(n) > 0$, which indicates that the orbit $k$ is not allowed to serve the task $i$ at the $n$-th iteration, the elements in $CO_i'(n)$ will be adjusted according to \textbf{Rules 2-4}. If $\sigma _{ik} = 0$, which means that the orbit $k$ is available to serve the task $i$ at the $n$-th iteration, $CO_i'(n)$ will be updated based on \textbf{Rule 5}. Finally, the tabu factor ${\sigma _{ik}}(n)$ will be updated iteratively according to \textbf{Rules 6-7}.

\textbf{Rule1:} In the initial situation, all the orbits in $C{O_i}$ can be used to serve the task $i$, which can be expressed as $CO_i'(1) = C{O_i}$.

\textbf{Rule2:} If there is only one visible orbit for the task $i$ (i.e., $\left| {C{O_i}} \right| = 1$), the orbit in $C{O_i}$ is always available, which can be expressed by $CO_i'(n) \equiv C{O_i}$. 

\textbf{Rule3:} If there are multiple visible orbits for the task $i$ (i.e., $\left| {C{O_i}} \right| > 1$), a randomly selected orbit $k$ would be removed from $CO_i'(n)$ to obtain  $CO_i'(n + 1)$, which is denoted as $CO_i'(n + 1) = CO_i'(n)/\left\{ {{O_k}} \right\}$.

\textbf{Rule4:} If there are multiple visible orbits and no orbits can be used for serving the task $i$ at the $n$-th iteration (i.e., $\left| {C{O_i}} \right| > 1$ and $CO_i'(n) = \emptyset $), $CO_i'(n)$ will be initialized as $CO_i'(n + 1) = C{O_i}$ .

\textbf{Rule5:} In the iterative process, when ${\sigma _{ik}}(n)$ decreases to 0, the orbit $k$ would be added to $CO_{i}'(n+1)$ in the next iteration, which is written as ${\kern 1pt} CO_i'(n + 1) = CO_i'(n) \cup {O_k}{\kern 1pt} {\kern 1pt} $.

\textbf{Rule6:} If the task $i$ allocated to the orbit $k$ is not successfully scheduled at the $n$-th iteration, the tabu factor ${\sigma _{ik}}(n)$ will be updated at the $(n+1)$-th iteration as ${\sigma _{ik}}(n+1) = l$, where $l\in \mathbb{Z}_{+}$.

\textbf{Rule7:} If the orbit $k$ is forbidden to serve the task $i$ at the $n$-th iteration (i.e., ${\sigma _{ik}}(n) > 0$), the tabu factor ${\sigma _{ik}}(n)$ will decrease according to ${\sigma _{ik}}(n + 1) = {\sigma _{ik}}(n){\kern 1pt}  - \Delta l$ in the following iterations until its value is 0. Here $\Delta l$ is a divisor of $l$, which is used to gradually reduce the value of ${\sigma _{ik}}(n)$.

\subsection{Ensemble of meta-heuristic and exact algorithm based on DCF}
The pseudo-code of EHE-DCF is provided in Algorithm \ref{al1}. In the algorithm, first, the task $i$ with the highest allocation priority is selected as the current task to be allocated (line 5). Second, according to the visibility of the orbits to the task $i$ and the orbit tabu condition, an available orbit set $CO_i'(n)$ is derived for serving task $i$ (line 6). Third, the allocation probabilities between the task $i$ and each orbit in $CO_i'(n)$ are calculated, and the task $i$ is allocated to an orbit $k$ based on the allocation probability ${p_{ik}}(n)$ (lines 8-9). Meanwhile, the pheromone is updated at each iteration (line 11). When all the tasks in $T$ are scheduled, the task allocation phase is terminated. 
In the single orbit scheduling phase, we use the CPLEX software to implement B\&B method to obtain a set of scheduling results of the single orbit scheduling problems. Then, the sub-scheduling results are merged to obtain an overall task scheduling result (lines 13-14). The task allocation and the single orbit scheduling phases are performed iteratively until the number of iterations reaches the maximum iterations $G1$.
\begin{algorithm}[ht]
    \caption{EHE-DCF}\label{al1}
    \KwIn{Task set $T$, orbit set $O$, time windows between tasks and orbits, maximum iterations $G$}
    \KwOut{scheduling results.}
    Initialize algorithm parameters\\
    \While(){$n \le G1$}{
        //\textit{Task allocation phase}\\
        \While(){$T \neq \emptyset$}{
            Find the task $i$ that has the highest allocation priority\\
            Select an available orbit set $CO_i'(n)$ for the task $i$\\
            \For(){each orbit $k \in CO_{i}'(n)$}{
                Calculate the allocation probability ${p_{ik}}(n)$ according to ${\tau _{ik}}(n)$ and ${\eta _{ik}}(n)$\\
            }
            Allocate the task $i$ to the orbit $k$ according to $p_{ik}(n)$\\
            Remove the task $i$ from $T$\\
            Update the pheromone\\
        }
        //\textit{Task scheduling phase}\\
        \For(){each orbit $k$ scheduled}{
            Solve the scheduling problem on the orbit $k$ by B\&B method\\
        }
        Merge all the sub-scheduling results;\\
        $n = n+1$\\ 
    }
   
\end{algorithm}

\subsection{Complexity analysis}
The computational complexity of each component of EHE-DCF is displayed in Table \ref{table:2}. Assume that the number of tasks allocated to each orbit varies in $[0, N]$. The number of tasks allocated to every single orbit will be $N$ in the worst case, and it would be much less than $N$ in practice. Assume that $N$ tasks are evenly allocated to $H$ orbits, the complexity of task allocation would be $O(N^2)$ and the complexity of single orbit scheduling would be $O(2^{N/H})$. Therefore, the complexity of EHE-DCF is $O(G*H*2^{N/H})$, where $G$ and $H$ are numbers of iterations and orbits, respectively. On the other hand, the complexity of the pure B\&B is $O(2^N)$ \cite{morrison2016branch}. Therefore, compared with the pure B\&B, the complexity of EHE-DCF is much smaller. As the resources and tasks scale increase, the computational advantage of EHE-DCF would become more obvious.
\begin{table}[htb]
\footnotesize
	\caption{The complexity of each component of EHE-DCF.}\label{table:2}
	\centering
	\begin{tabular}{ll}
	\hline
	Components & Complexity \\
	\hline
    Initialization &$O(1)$\\
    Task allocation &$O(N^2)$\\
    Single orbit scheduling &$O(2^{N/H})$\\
	\hline
	\end{tabular}
\end{table}
\section{Simulation experiments}\label{s6}
In this section, we carry out experiments based on EOS scheduling instances with different task scales and different observation resource scales to comprehensively evaluate the performance of EHE-DCF. Both EHE-DCF and its competitors are implemented in Matlab R2016a and CPLEX12.5, and all the experiments are executed on a computer with Intel(R) Core (TM) i5 2.80 GHz and 8.0 GB RAM.

\subsection{Comparative algorithms}
We compare EHE-DCF with five algorithms, including three DCF based meta-heuristics, a state-of-the-art meta-heuristic in the existing literature, and pure B\&B method. The pure B\&B method is implemented with the commercial solver CPLEX, and the meta-heuristics are briefly introduced as follows.

(1) \textit{Greedy algorithm based on DCF (GR-DCF)}. Greedy algorithms that preferentially schedule the task with the highest profit or priority are commonly used to solve satellite scheduling problems in practical applications \cite{CHEN2018177,WU2019taskpriority}. In GR-DCF, the task allocation phase is the same as EHE-DCF, while the task scheduling phase is performed greedily. In each iteration of the task scheduling phase, the tasks assigned to each orbit are scheduled iteratively according to their profits. During the scheduling process, if constraints (\ref{eq3})-(\ref{eq4}) are violated when a task is inserted into an orbit, the previously scheduled tasks on this orbit would be removed one by one, until all constraints are satisfied. The scheduled task with the lowest profit would be removed first. Finally, the tasks that are not successfully scheduled are preserved for the next iteration.

(2) \textit{Simulated annealing neighborhood based on greedy neighborhood structure and DCF (SANS1-DCF)}. Different from GR-DCF, SANS1-DCF implements a greedy neighborhood structure in the task scheduling phase. The greedy neighborhood structure schedules tasks assigned to each orbit in the same way as GR-DCF, but removes a task with the lowest profit before inserting tasks if any task has been inserted into the orbit. By removing tasks from previously scheduled results, SANS1-DCF is expected to have a higher capability to escape from the local optimum. Meanwhile, SANS1-DCF adopts the well-known Metropolis acceptance criteria \cite{Peng2011simulated} of simulated annealing algorithm to accept worse solutions with an adaptively controlled probability.

(3) \textit{Simulated annealing neighborhood based on random neighborhood structure and DCF (SANS2-DCF)}. The framework of SANS2-DCF is similar with SANS1-DCF. The difference between SANS1-DCF and SANS2-DCF is that SANS2-DCF removes a task randomly instead of removing the task with the lowest profit from the previously scheduled result.

(4) \textit{Adaptive simulated annealing–based scheduling algorithm (ASA)}. ASA is a state-of-the-art algorithm extracted from the existing literature \cite{WU2017576}. This algorithm has been proved efficient in solving EOS scheduling problems, due to involving sophisticated mechanisms, i.e., adaptive temperature control, tabu-list-based short-term revisiting avoidance mechanism and intelligent combination of neighbourhood structures.

\subsection{Experiment setups}
In the experimental studies, 8 instances varying from 200 to 1600 tasks are prepared. The targets associated with the tasks are distributed in a range of latitude 15°-45° and longitude 80°-120° randomly. The profits of tasks are random values within [1, 10]. We set the allowable runtime for an algorithm solving a scheduling problem to 3600 seconds and the scheduling horizon to 24 hours. The basic information of the instances is provided in Table \ref{table:3}. Parameters of satellites and algorithms are listed in Tables \ref{table:4} and \ref{table:5}, respectively. All algorithms are repeated 25 times on each instance independently.
\begin{table}[h]
\footnotesize
	\caption{Information of instances.}\label{table:3}
	\centering
	\begin{tabular}{lllll}
	\hline
	Scheduling horizon& Number of satellites& Number of tasks& Profit of task& Acceptable runtime\\
    \hline
    24 hours& 10& [200, 1600]& [1,10]& 3600 seconds\\
	\hline
	\end{tabular}
\end{table}
\begin{table}[htb]
\footnotesize
	\caption{Parameters of satellites.}\label{table:4}
	\centering
    \footnotesize
	\begin{tabular}{llllllll}
	\hline
    $E_k$& $M_k$& $e_k$& $m_k$& $td_k$& $ts_k$& $tu_k$& $v$\\
    \hline
    300& 2400& 1& 1& 5& 3& 5& 1\\
	\hline
	\end{tabular}
\end{table}
\begin{table}[htb]
\footnotesize
	\caption{Parameters of algorithms.}\label{table:5}
	\centering
	\begin{tabular}{lp{0.7\textwidth}}
	\hline
    Algorithms& Parameters\\
    \hline
    EHE-DCF& $\alpha=3$, $\beta=3$, $a=0.7$, $b=0.3$, $c=0.2$, $l=2$, $\Delta l=1$, $\rho=0.1$, $G=200$.\\
    \hline
    GR-DCF& \multirow{3}{*}{\shortstack[l]{Maximum iterations $G2=500$, \\start temperature $Tem_s=300$,\\ end temperature $Tem_e=0.001$,\\ cooling rate $\delta=0.99$.}}\\
    SANS1-DCF& \\
    SANS2-DCF& \\
	\hline
    ASA& Maximum iterations $G3=5*N$ and the other parameters are the same as in \cite{WU2017576}.\\
    \hline
	\end{tabular}
\end{table}

\subsection{Results and discussions}
The results are reported in Table \ref{table:6}, including the obtained observation profits, number of scheduled tasks, and average runtime. As seen from Table \ref{table:6}, EHE-DCF outperforms its comparative meta-heuristics (i.e., GR-DCF, SANS1-DCF, SANS2-DCF, and ASA) in terms of the obtained observation profits and the number of scheduled tasks. This is because EHE-DCF uses B\&B method to generate optimal solutions for single orbit scheduling sub-problems and the iterative task allocation procedure realizes a proper problem partition. Meanwhile, these two phases can work cooperatively to obtain a high-quality entire scheduling scheme. 

By contrast, pure B\&B can get the highest profits in the small-scale task scheduling problems without surprise, as pure B\&B is an exact algorithm. However, its runtime increases dramatically when the task scale increases, indicating that its computational efficiency is not satisfactory when solving large-scale EOS scheduling problems. To be concrete, pure B\&B has a sharp increase in runtime when the number of tasks is more than 600, and it consumes much more time than other comparative algorithms. In the case of 800 tasks, the runtime of pure B\&B has exceeded the predefined allowable running time. As for ASA, it needs more computational efforts to solve the scheduling problem, while the obtained profits are less than that of other meta-heuristics based on DCF. Particularly, ASA cannot solve large-scale instances (i.e., instances C5-C8) within acceptable running time. Although pure B\&B consumes less computational time when solving small-scale instances (i.e., instances C1-C3) compared with ASA, the pure B\&B still requires more computational time compared with DCF based meta-heuristics on instances C2-C8. These observation results further prove the efficiency of DCF, particularly on large-scale instances.
\begin{table}[htp]
    \caption{Experimental results on 8 instances.}\label{table:6}
    \centering
    \resizebox{\textwidth}{!}{%
    \begin{tabular}{cccp{0.1\textwidth}p{0.1\textwidth}p{0.1\textwidth}p{0.1\textwidth}p{0.1\textwidth}p{0.1\textwidth}c}
        \hline
        &                              &                                 & \multicolumn{3}{c}{Profits}                                                                   & \multicolumn{3}{c}{Number of scheduled tasks}                                           &                                 \\ \cline{4-9}
\multirow{-2}{*}{Instance} & \multirow{-2}{*}{Task scale} & \multirow{-2}{*}{Algorithm}     & Min.                        & Ave.                            & Max.                         & Min.                        & Ave.                        & Max.                        & \multirow{-2}{*}{Runtime (s)}   \\ \hline
        &                              & \cellcolor[HTML]{C0C0C0}EHE-DCF & \cellcolor[HTML]{C0C0C0}1063 & \cellcolor[HTML]{C0C0C0}1095.84 & \cellcolor[HTML]{C0C0C0}1117 & \cellcolor[HTML]{C0C0C0}160 & \cellcolor[HTML]{C0C0C0}166 & \cellcolor[HTML]{C0C0C0}170 & \cellcolor[HTML]{C0C0C0}16.131  \\
        &                              & SANS1-DCF                       & 1046                         & 1073.88                         & 1091                         & 157                         & 162                         & 167                         & 10.297                          \\
        &                              & SANS2-DCF                       & 1054                         & 1077.88                         & 1102                         & 157                         & 163                         & 167                         & 9.056                           \\
        &                              & GR-DCF                          & 1036                         & 1060.52                         & 1084                         & 157                         & 162                         & 168                         & 4.734                           \\
        &                              & Pure B\&B                       & \textbackslash{}             & 1198.00                            & \textbackslash{}             & \textbackslash{}            & 186                         & \textbackslash{}            & 3.104                           \\
\multirow{-6}{*}{C1}       & \multirow{-6}{*}{200}        & ASA                             & 1050                         & 1069.90                          & 1098                         & 161                         & 164                         & 168                         & 37.866                          \\ \hline
        &                              & \cellcolor[HTML]{C0C0C0}EHE-DCF & \cellcolor[HTML]{C0C0C0}1833 & \cellcolor[HTML]{C0C0C0}1859.48 & \cellcolor[HTML]{C0C0C0}1902 & \cellcolor[HTML]{C0C0C0}253 & \cellcolor[HTML]{C0C0C0}259 & \cellcolor[HTML]{C0C0C0}266 & \cellcolor[HTML]{C0C0C0}27.744  \\
        &                              & SANS1-DCF                       & 1750                         & 1781.28                         & 1848                         & 242                         & 247                         & 253                         & 44.782                          \\
        &                              & SANS2-DCF                       & 1773                         & 1807.12                         & 1834                         & 242                         & 249                         & 252                         & 41.367                          \\
        &                              & GR-DCF                          & 1654                         & 1728.92                         & 1781                         & 237                         & 246                         & 250                         & 18.488                          \\
        &                              & Pure B\&B                       & \textbackslash{}             & 2134.00                            & \textbackslash{}             & \textbackslash{}            & 306                         & \textbackslash{}            & 240.753                         \\
\multirow{-6}{*}{C2}       & \multirow{-6}{*}{400}        & ASA                              & 1541                         & 1591.00                            & 1653                         & 215                         & 223                         & 231                         & 432.491                         \\ \hline
        &                              & \cellcolor[HTML]{C0C0C0}EHE-DCF & \cellcolor[HTML]{C0C0C0}2235 & \cellcolor[HTML]{C0C0C0}2280.64 & \cellcolor[HTML]{C0C0C0}2312 & \cellcolor[HTML]{C0C0C0}293 & \cellcolor[HTML]{C0C0C0}302 & \cellcolor[HTML]{C0C0C0}308 & \cellcolor[HTML]{C0C0C0}54.801  \\
        &                              & SANS1-DCF                       & 2078                         & 2133.64                         & 2181                         & 273                         & 280                         & 287                         & 153.926                         \\
        &                              & SANS2-DCF                       & 2149                         & 2208.36                         & 2252                         & 277                         & 287                         & 297                         & 147.857                         \\
        &                              & GR-DCF                          & 1974                         & 2042.08                         & 2100                         & 270                         & 277                         & 287                         & 62.184                          \\
        &                              & Pure B\&B                       & \textbackslash{}             & 2616.00                            & \textbackslash{}             & \textbackslash{}            & 350                         & \textbackslash{}            & 448.514                         \\
\multirow{-6}{*}{C3}       & \multirow{-6}{*}{600}        & ASA                              & 1703                         & 1744.00                            & 1801                         & 231                         & 237                         & 243                         & 1583.542                        \\ \hline
        &                              & \cellcolor[HTML]{C0C0C0}EHE-DCF & \cellcolor[HTML]{C0C0C0}2531 & \cellcolor[HTML]{C0C0C0}2577.92 & \cellcolor[HTML]{C0C0C0}2616 & \cellcolor[HTML]{C0C0C0}318 & \cellcolor[HTML]{C0C0C0}327 & \cellcolor[HTML]{C0C0C0}335 & \cellcolor[HTML]{C0C0C0}102.31  \\
        &                              & SANS1-DCF                       & 2306                         & 2358.92                         & 2420                         & 287                         & 296                         & 305                         & 383.494                         \\
        &                              & SANS2-DCF                       & 2407                         & 2465.12                         & 2521                         & 298                         & 307                         & 314                         & 376.98                          \\
        &                              & GR-DCF                          & 2194                         & 2255.76                         & 2316                         & 286                         & 295                         & 309                         & 159.789                         \\
        &                              & Pure B\&B                       & \textbackslash{}             & 2928.00                            & \textbackslash{}             & \textbackslash{}            & 371                         & \textbackslash{}            & 4472.515                        \\
\multirow{-6}{*}{C4}       & \multirow{-6}{*}{800}        & ASA                              & 1877                         & 1923.90                          & 1976                         & 238                         & 247                         & 255                         & 3519.338                        \\ \hline
        &                              & \cellcolor[HTML]{C0C0C0}EHE-DCF & \cellcolor[HTML]{C0C0C0}2762 & \cellcolor[HTML]{C0C0C0}2796.36 & \cellcolor[HTML]{C0C0C0}2829 & \cellcolor[HTML]{C0C0C0}341 & \cellcolor[HTML]{C0C0C0}346 & \cellcolor[HTML]{C0C0C0}354 & \cellcolor[HTML]{C0C0C0}170.073 \\
        &                              & SANS1-DCF                       & 2430                         & 2530.28                         & 2588                         & 303                         & 310                         & 315                         & 756.406                         \\
        &                              & SANS2-DCF                       & 2544                         & 2621.84                         & 2674                         & 307                         & 317                         & 324                         & 767.229                         \\
\multirow{-4}{*}{C5}       & \multirow{-4}{*}{1000}       & GR-DCF                          & 2371                         & 2403.04                         & 2467                         & 301                         & 307                         & 313                         & 322.98                          \\ \hline
        &                              & \cellcolor[HTML]{C0C0C0}EHE-DCF & \cellcolor[HTML]{C0C0C0}2903 & \cellcolor[HTML]{C0C0C0}2936.04 & \cellcolor[HTML]{C0C0C0}2976 & \cellcolor[HTML]{C0C0C0}354 & \cellcolor[HTML]{C0C0C0}358 & \cellcolor[HTML]{C0C0C0}365 & \cellcolor[HTML]{C0C0C0}274.739 \\
        &                              & SANS1-DCF                       & 2576                         & 2628.68                         & 2684                         & 306                         & 315                         & 326                         & 1357.924                        \\
        &                              & SANS2-DCF                       & 2729                         & 2769.96                         & 2857                         & 323                         & 329                         & 338                         & 1338.86                         \\
\multirow{-4}{*}{C6}       & \multirow{-4}{*}{1200}       & GR-DCF                          & 2407                         & 2483.60                          & 2562                         & 306                         & 313                         & 320                         & 574.383                         \\ \hline
        &                              & \cellcolor[HTML]{C0C0C0}EHE-DCF & \cellcolor[HTML]{C0C0C0}3036 & \cellcolor[HTML]{C0C0C0}3074.88 & \cellcolor[HTML]{C0C0C0}3108 & \cellcolor[HTML]{C0C0C0}361 & \cellcolor[HTML]{C0C0C0}368 & \cellcolor[HTML]{C0C0C0}372 & \cellcolor[HTML]{C0C0C0}422.069 \\
        &                              & SANS1-DCF                       & 2676                         & 2731.80                          & 2795                         & 314                         & 322                         & 331                         & 2178.727                        \\
        &                              & SANS2-DCF                       & 2818                         & 2883.56                         & 2943                         & 328                         & 337                         & 344                         & 2183.903                        \\
\multirow{-4}{*}{C7}       & \multirow{-4}{*}{1400}       & GR-DCF                          & 2509                         & 2577.417                        & 2680                         & 310                         & 319                         & 330                         & 922.881                         \\ \hline
        &                              & \cellcolor[HTML]{C0C0C0}EHE-DCF & \cellcolor[HTML]{C0C0C0}3108 & \cellcolor[HTML]{C0C0C0}3149.08 & \cellcolor[HTML]{C0C0C0}3187 & \cellcolor[HTML]{C0C0C0}368 & \cellcolor[HTML]{C0C0C0}374 & \cellcolor[HTML]{C0C0C0}379 & \cellcolor[HTML]{C0C0C0}620.367 \\
        &                              & SANS1-DCF                       & 2738                         & 2789.16                         & 2831                         & 320                         & 326                         & 333                         & 3354.379                        \\
        &                              & SANS2-DCF                       & 2883                         & 2945.48                         & 3006                         & 330                         & 341                         & 350                         & 3318.207                        \\
\multirow{-4}{*}{C8}       & \multirow{-4}{*}{1600}       & GR-DCF                          & 2555                         & 2634.80                          & 2693                         & 314                         & 322                         & 333                         & 1396.34                         \\ \hline
        &                              &                                 &                              &                                 &                              &                             &                             &                             &                                
\end{tabular}
    }
\end{table}

\begin{figure}[!htb]
	\centering
	\includegraphics[width=0.5\textwidth]{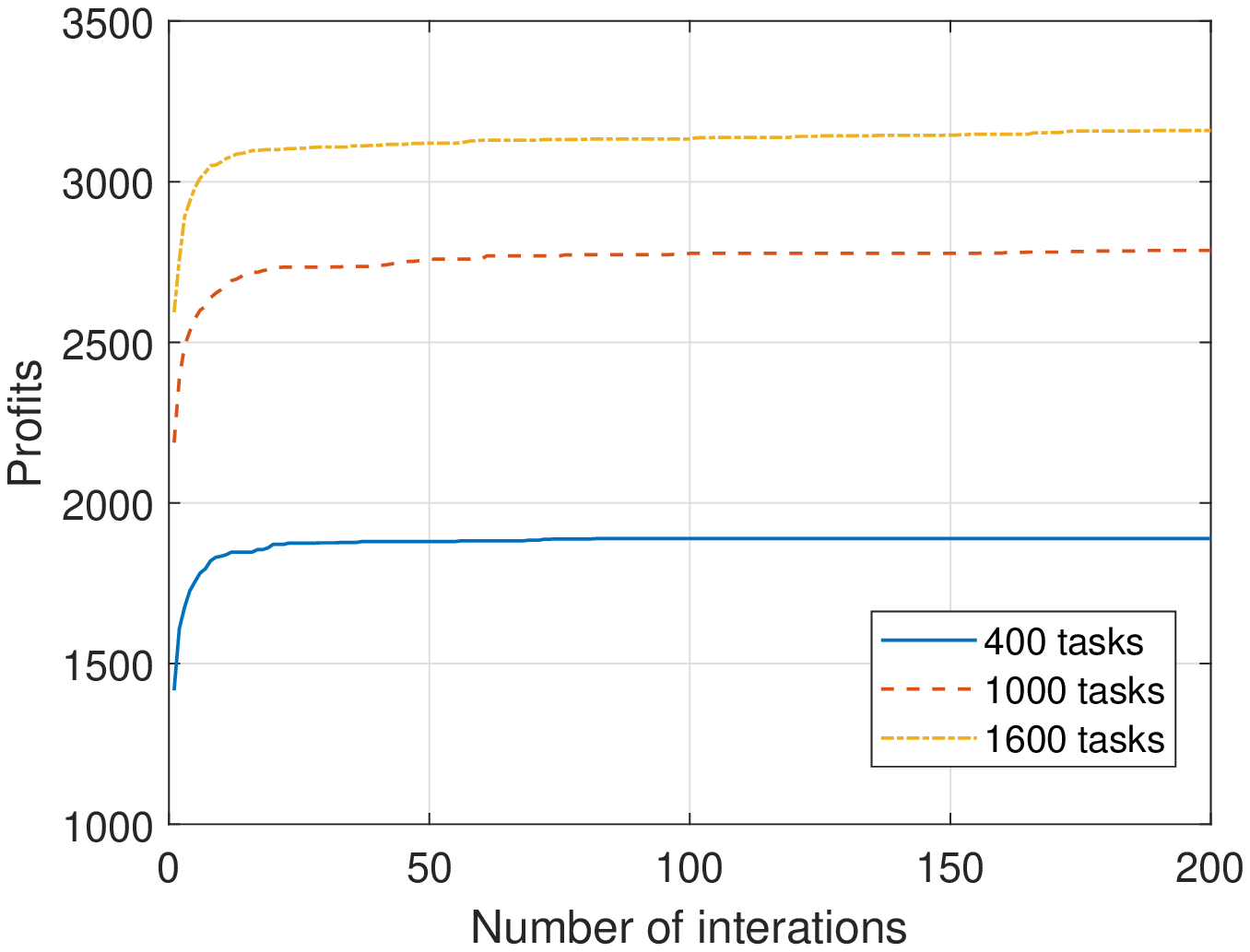}
	\caption{Convergence of EHE-DCF on the instances with 400, 1000, and 1600 tasks.}
	\label{Fig:5}
\end{figure}

To visually analysis the performance of EHE-DCF, we plot the convergence process of EHE-DCF when dealing with the instances with 400, 1000, and 1600 tasks in Figure \ref{Fig:5}, which demonstrates that EHE-DCF is robust and able to converge to a satisfactory solution efficiently. 

\begin{figure}[!htb]
	\centering
	\subfigure[Variance of the obtained profits.]{
	\includegraphics[width = 0.8\textwidth]{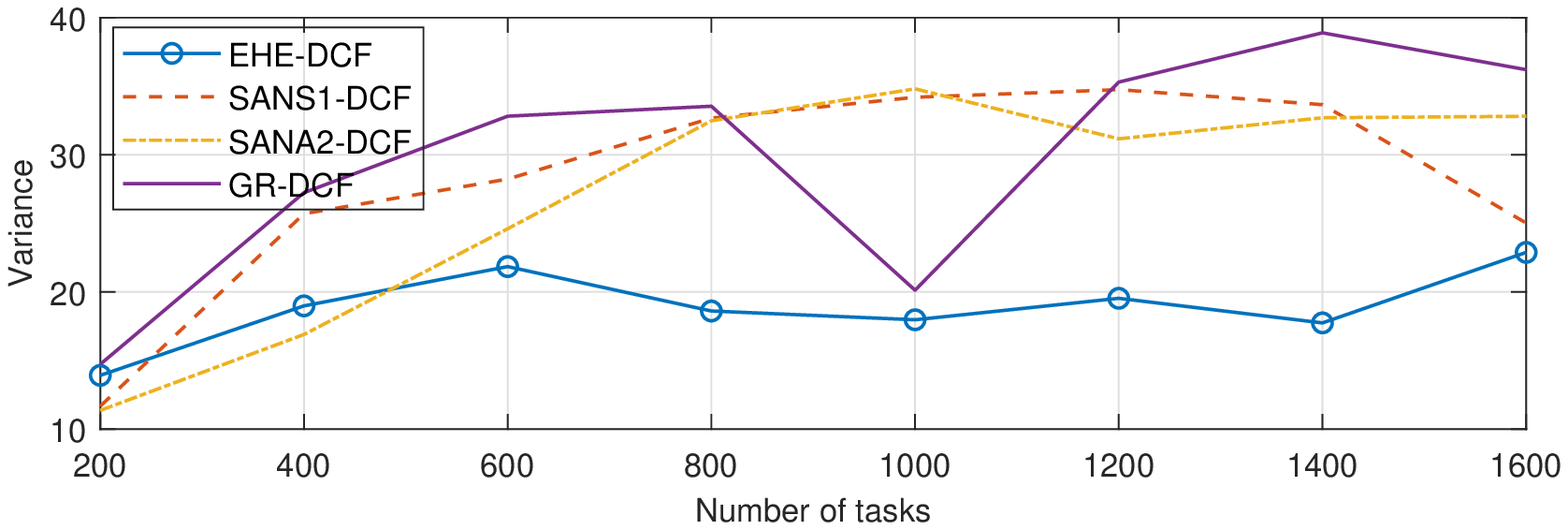}
	}\\
	\subfigure[Variance of the number of scheduled tasks.]{
	\includegraphics[width = 0.8\textwidth]{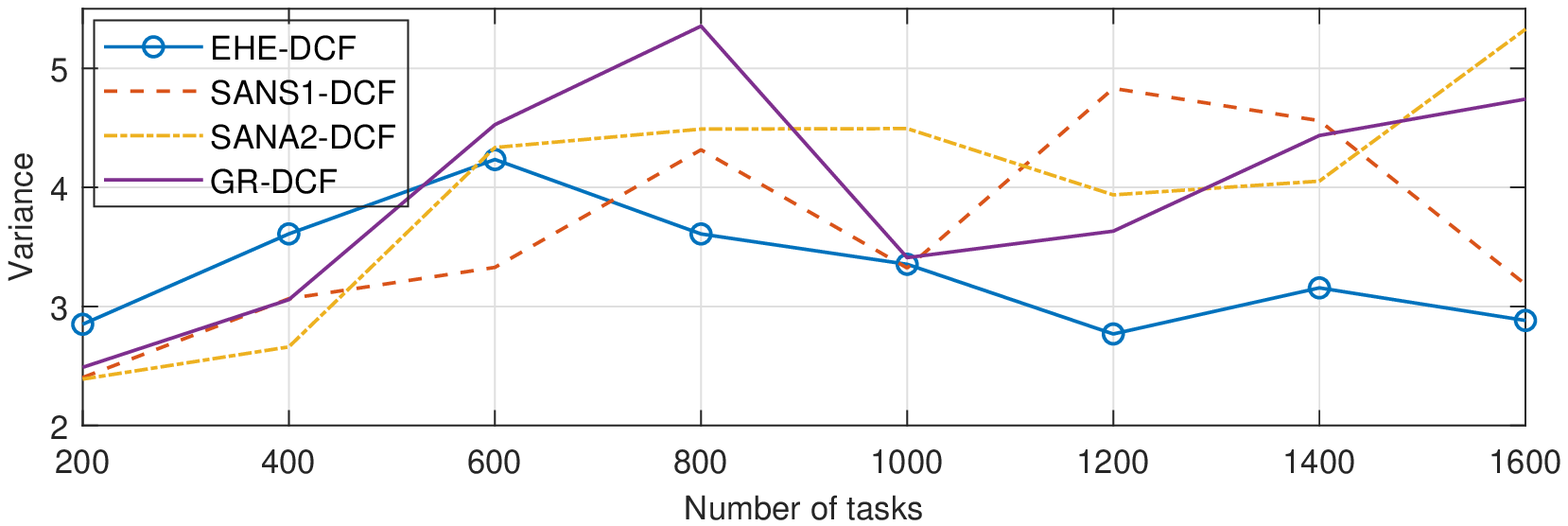}
	}
	\caption{Comparisons on the stability with respect to the obtained profits and the number of scheduled tasks.}
	\label{Fig:6}
\end{figure}

The variances of the scheduling profits and number of scheduled tasks obtained by four DCF based meta-heuristics on all instances are investigated in Figure \ref{Fig:6}. It can be found that EHE-DCF has a more stable performance in solving EOS scheduling problems with different scales compared with SANS1-DCF, SANS2-DCF, and GR-DCF. In particular, its advantage is more obvious when the task scale is more than 800, which indicates that EHE-DCF could be useful and reliable in practical applications.

\begin{figure}[!htb]
	\centering
	\subfigure[Increase rate of the obtained profits.]{
	\includegraphics[width = 0.45\textwidth]{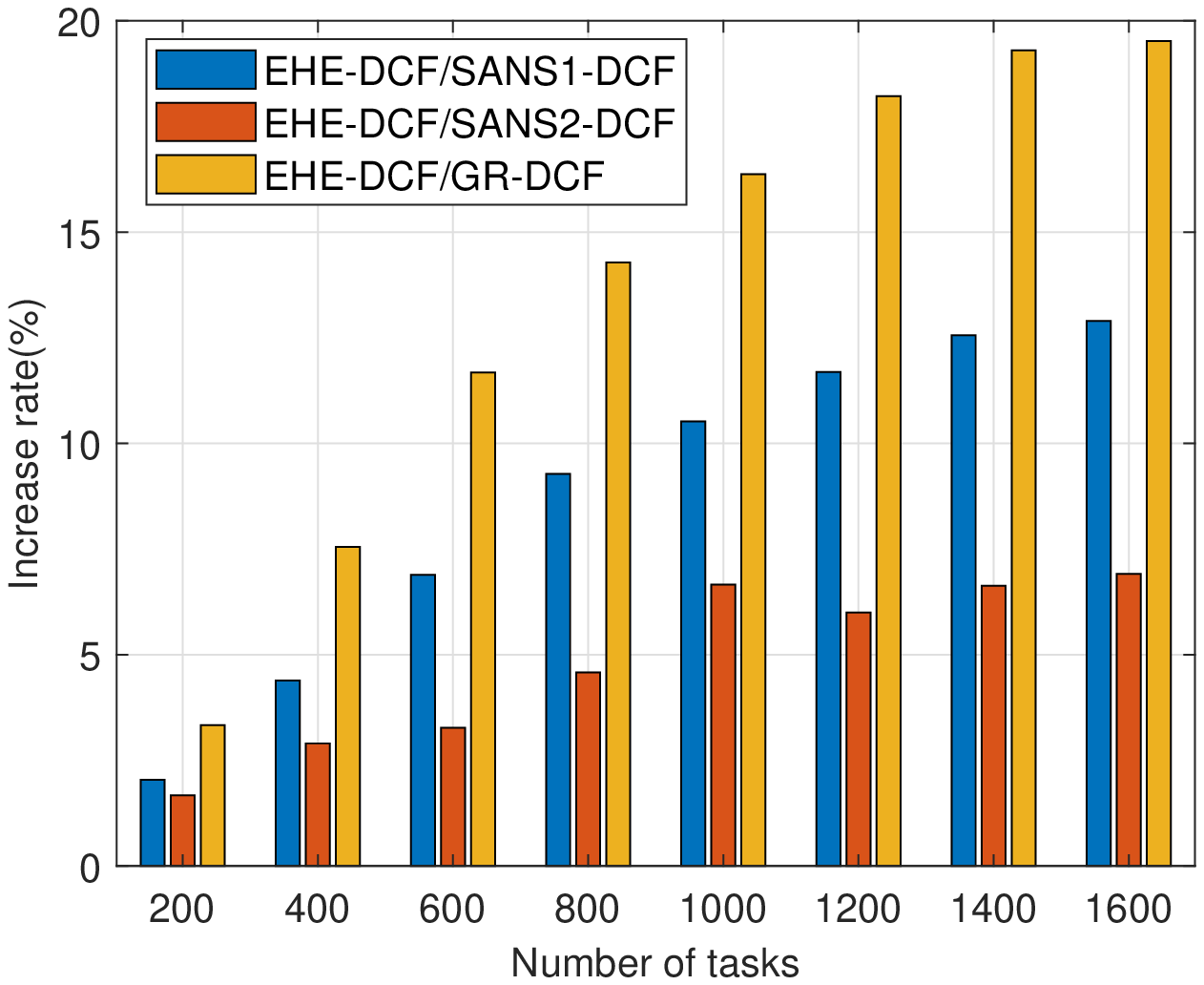}
	}
	\subfigure[Increase rate of the number of scheduled tasks.]{
	\includegraphics[width = 0.45\textwidth]{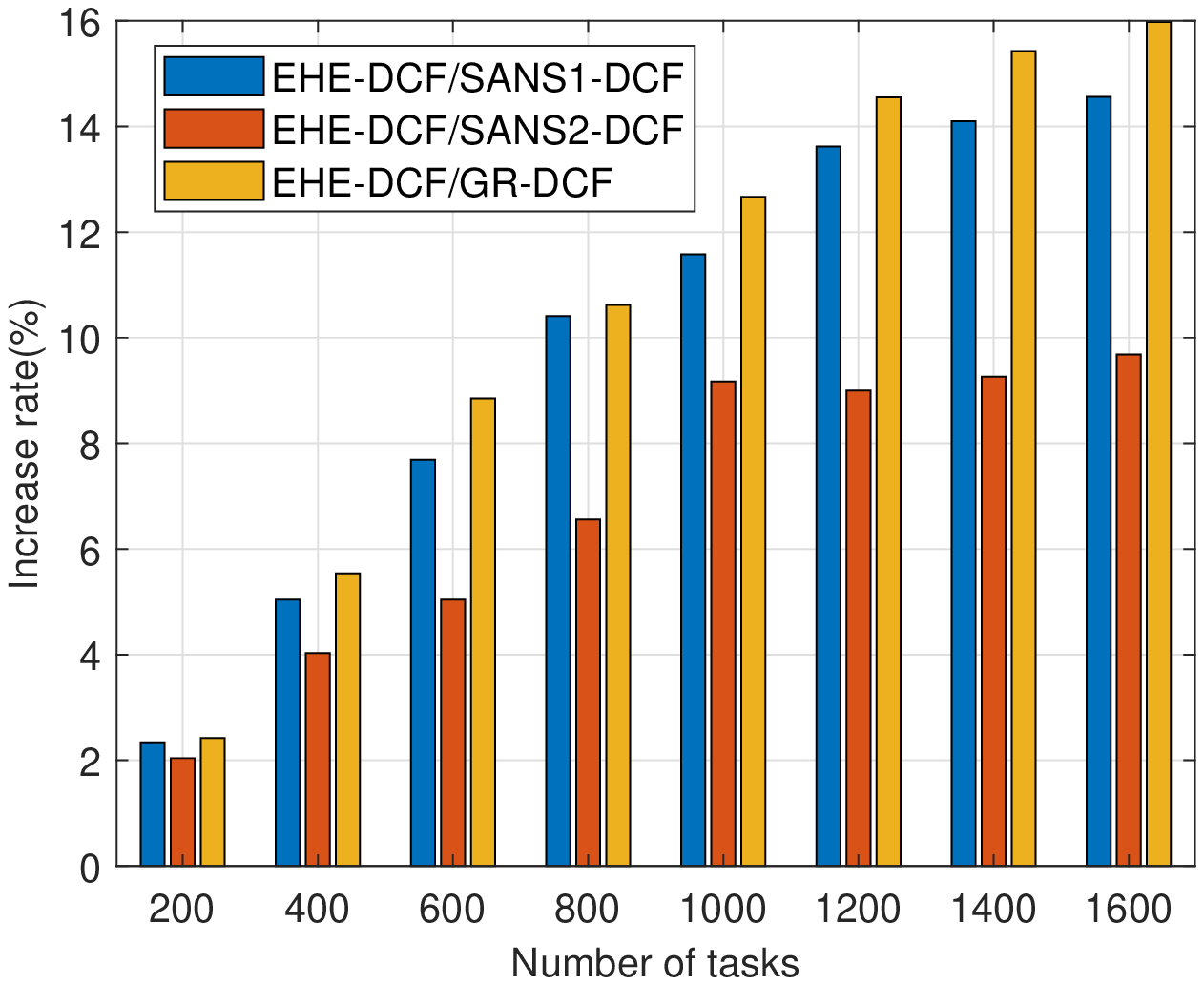}
	}
	\caption{Histograms on different task scales with respect to the obtained profits and the number of scheduled tasks.}
	\label{Fig:7}
\end{figure}

To further demonstrate the advantages of EHE-DCF under large-scale task scheduling, we define two indicators, i.e., the increase rate of obtained profits $r_{1,i}^{profit}$ and the number of scheduled tasks $r_{1,i}^{task}$, which are respectively calculated as follows
\begin{equation}\label{eq24}
    r_{1,i}^{profit}=(\gamma_{1}^{ave}-\gamma_{i}^{ave})/\gamma_{i}^{ave}\times100\%, i=\{2,3,4\},
\end{equation}
\begin{equation}\label{eq25}
    r_{1,i}^{task}=(Num{1}^{ave}-Num{i}^{ave})/Num_{i}^{ave}\times100\%, i=\{2,3,4\}.
\end{equation}
where $\gamma_{i}^{ave}$ and $Num{i}^{ave}$ represent the average value of obtained profits and the number of scheduled tasks of the DCF based algorithm $i$, respectively. The comparison results of increase rates are shown in Figure \ref{Fig:7}. The results show that when the task scale is 1600, the increase rates of the obtained profits and number of scheduled tasks reach the maximum values. It can be concluded that the performance of three comparative DCF based meta-heuristics (i.e., GR-DCF, SANS1-DCF, and SANS2-DCF) are deteriorated when solving large-scale EOS scheduling problems. On the contrary, the increase rates of EHE-DCF increase significantly in the case of large-scale tasks, indicating that EHE-DCF is particularly suitable to large-scale complex EOS scheduling problems.

\begin{figure}[!htb]
	\centering
	\subfigure[Profits obtained by satellites with different scales.]{
	\includegraphics[width = 0.45\textwidth]{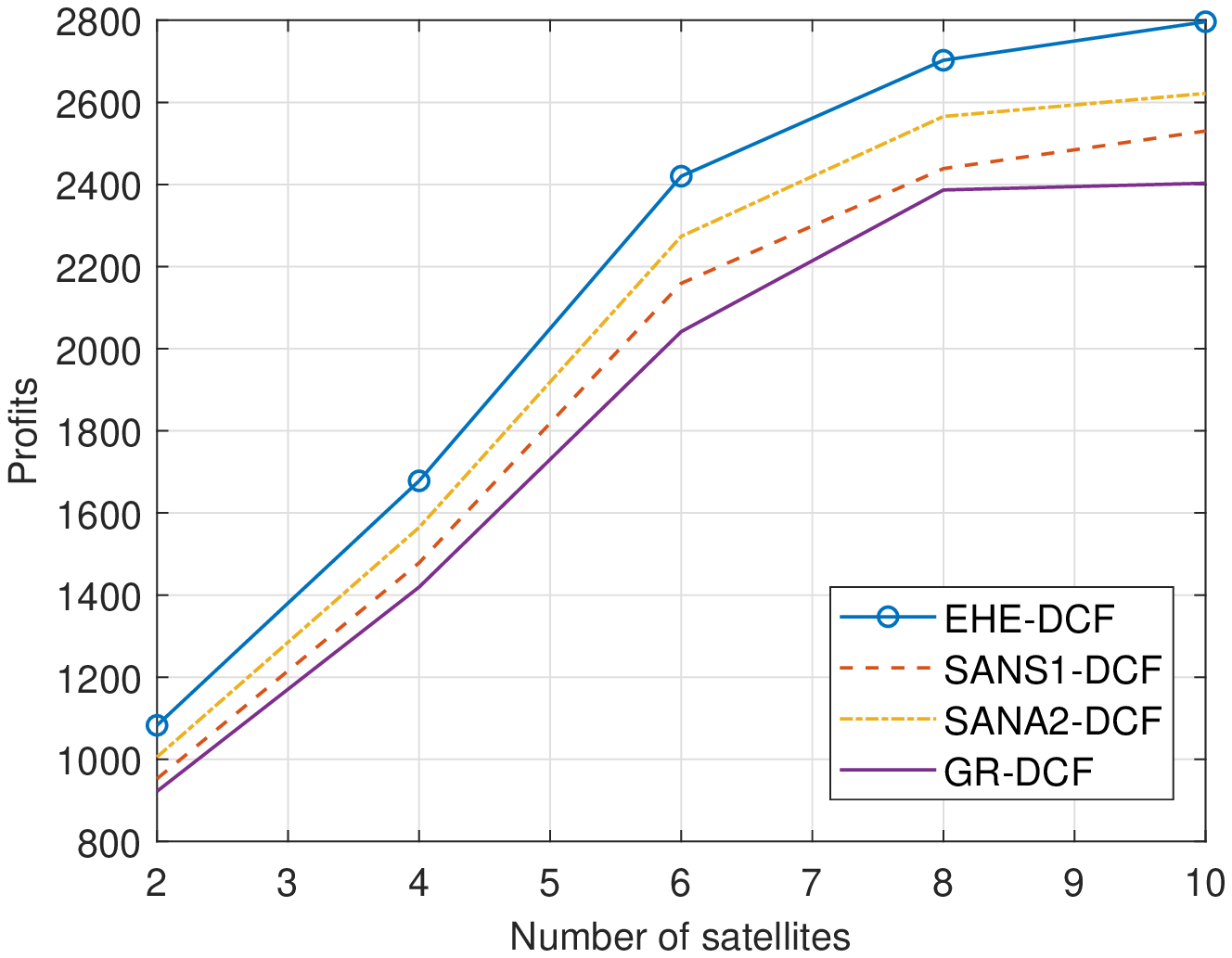}}
	\subfigure[Number of tasks scheduled by satellites with different scales.]{
	\includegraphics[width = 0.45\textwidth]{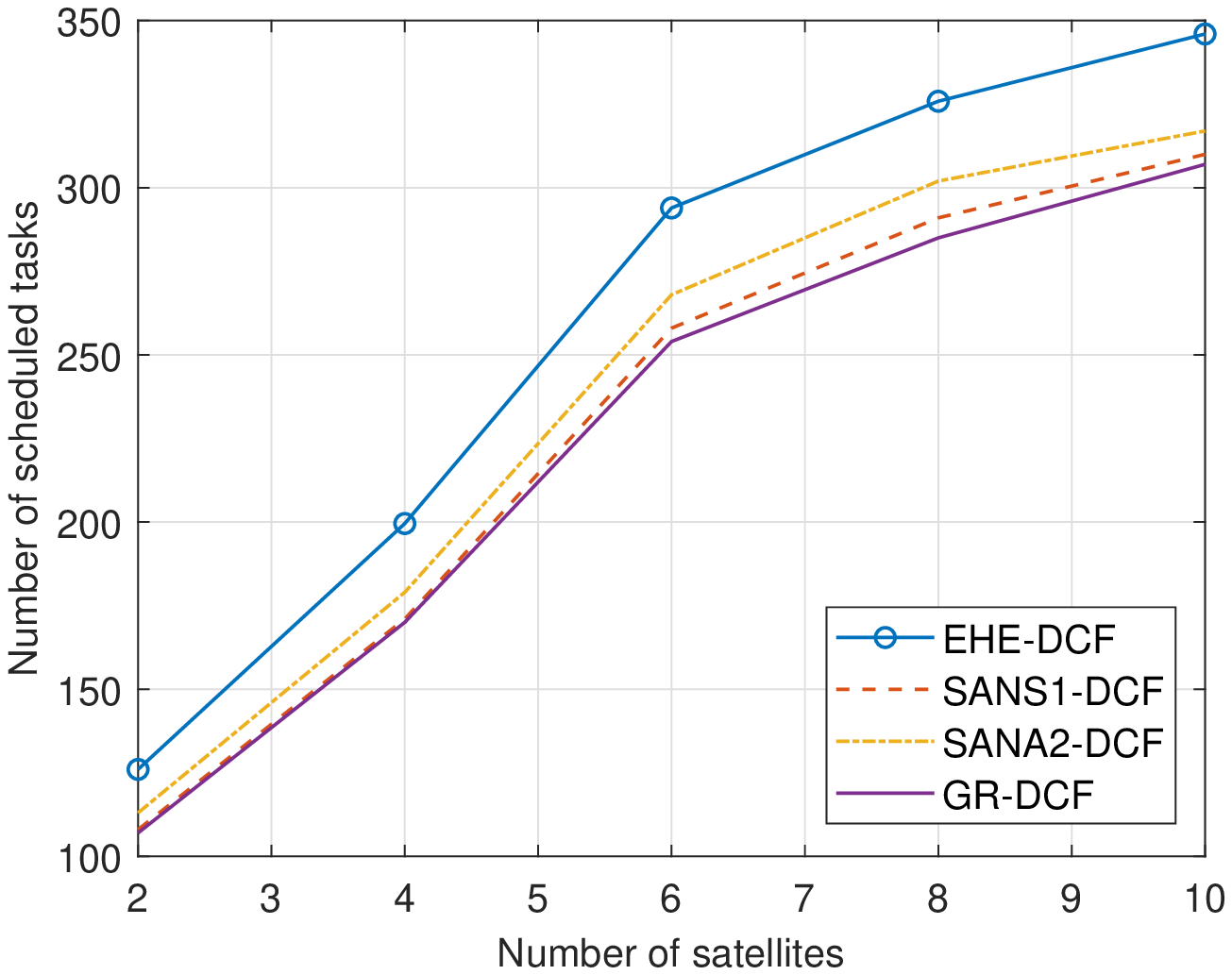}}
	\caption{Comparisons on different satellite scales with respect to the obtained profits and the number of scheduled tasks.}
	\label{Fig:8}
\end{figure}

More observation resources mean more observation opportunities, while it brings more scheduling challenges. To test the performance of EHE-DCF in solving EOS scheduling problems with different numbers of observation resources, we apply EHE-DCF, GR-DCF, SANS1-DCF, and SANS2-DCF to instances with a different number of satellites and the same number of tasks. Five groups of instances are implemented by setting the number of satellites from 2 to 10 and the number of tasks to 1000. ASA is not tested here for comparisons, as we have shown the superior performance of the DCF based meta-heuristics in Table~\ref{table:6}. The computational results based on EHE-DCF, SANS1-DCF, SANS2-DCF, and GR-DCF are plotted in Figure \ref{Fig:8}. It can be seen that the number of satellites shows a significant effect on the results of scheduling. Meanwhile, the scheduling results obtained by EHE-DCF are significantly better than those of other comparative algorithms on all instances.

\section{Conclusions}\label{s7}
In this paper, a novel ensemble approach named ECE-DCF, which combines meta-heuristic and exact methods based on a DCF, has been proposed for solving the multiple EOS scheduling problem. ECE-DCF divides the EOS scheduling problem into a task allocation phase and a single orbit scheduling phase. In the task allocation phase, a meta-heuristic is designed to generate a fairly reasonable task allocation scheme in an iterative manner. This meta-heuristic method involves sophisticated mechanisms, i.e., probabilistic selection and tabu mechanism, feedback factors, heuristic factors, and pheromone trail factors. In the single orbit scheduling phase, we construct an integer programming model and adopt the B\&B method to obtain an optimal solution of each subproblem. Furthermore, these two phases are performed iteratively and interactively to solve the EOS scheduling problem. Compared with an exact method (i.e., pure B\&B), three DCF based meta-heuristic (i.e., GR-DCF, SANS1-DCF and SANS2-DCF), and a state-of-the-art meta-heuristic (i.e., ASA), EHE-DCF outperforms the competitors in terms of scheduling profits and  number of scheduled tasks on the most instances, as well as running time on large-scale instances. Extensive experiments are further conducted to demonstrate that EHE-DCF is a robust and efficient method for solving EOS scheduling problems, especially when the scale of the scheduling problem getting large. In future studies, we would extend the proposed approach to solve more complicated EOS scheduling problems, scuh as the agile EOS scheduling problem \cite{wang2020agile}.

\section*{Acknowledgment}
This work was supported in part by the National Natural Science Foundation of China under Grants 61603404 and 71801218, in part by Natural Science Fund for Distinguished Young Scholars of Hunan Province under Grant 2019JJ20026.

\bibliographystyle{IEEEtran}
\bibliography{IEEEabrv,mybibfile}





\end{document}